\documentclass[aps,twocolumn,prd,showpacs,amsmath,superscriptaddress,floatfix]{revtex4}
\usepackage{txfonts}
\usepackage{graphicx}
\usepackage{bm}

\newcommand\Ln[1]{{\rm Ln}_{#1}}
\newcommand\Ex[1]{{\rm Ex}_{#1}}

\newcommand\apjs{\apj Suppl.\ Ser.}
\newcommand\aj{Astron.\ J.}
\newcommand\mnras{Mon.\ Not.\ R. Astron.\ Soc.}
\newcommand\physrep{Phys.\ Rep.}

\newcommand\beq{\begin{equation}}
\newcommand\eeq{\end{equation}}

\newcommand\bdm{\begin{displaymath}}
\newcommand\edm{\end{displaymath}}

\begin{document}

\title{Distribution function of the dark matter}

\author{N. Wyn Evans}\email{nwe@ast.cam.ac.uk}
\affiliation{Institute of Astronomy, University of Cambridge,
Madingley Road, Cambridge, CB3 0HA, United Kingdom}

\author{Jin H. An}\email{jinan@space.mit.edu}
\affiliation{Institute of Astronomy, University of Cambridge,
Madingley Road, Cambridge, CB3 0HA, United Kingdom}
\affiliation{MIT Kavli Institute for Astrophysics \& Space Research,
Massachusetts Institute of Technology,
77 Massachusetts Avenue, Cambridge, MA 02139, USA}

\begin{abstract}
There is good evidence from N-body simulations that the velocity
distribution in the outer parts of halos is radially anisotropic, with
the kinetic energy in the radial direction roughly equal to the sum of
that in the two tangential directions. We provide a simple algorithm
to generate such cosmologically important distribution functions.
Introducing $r_E(E)$, the radius of the largest orbit of a particle
with energy $E$, we show how to write down almost trivially a
distribution function of the form $f(E,L)=L^{-1}g(r_E)$ for any
spherical model -- including the `universal' halo density law
(Navarro-Frenk-White
profile). We in addition give the generic form of the distribution
function for any model with a local density power-law index $\alpha$
and anisotropy parameter $\beta$ and provide limiting forms
appropriate for the central parts and envelopes of dark matter halos.
From those, we argue that, regardless of the anisotropy, the density
falloff at large radii must evolve to $\rho\sim r^{-4}$ or steeper
ultimately.
\end{abstract}
 
\maketitle

\section{Introduction}

N-body experiments now can reliably follow the collapse and violent
relaxation of dark matter halos from initial conditions. This has led
to the discovery of regularities in the phase space distribution of
dark matter \citep[e.g.,][]{TN01}, even though the final state is not
completely independent of initial conditions. This is important
because it suggests that there is a generic functional form for the
distribution function (DF) that describes the physics of violent
relaxation, albeit with some cosmic scatter \citep{HMZ05}.

For example, \citet[see also \citealp{HS05}]{HM05} have found that the
density power index is correlated with the anisotropy parameter
$\beta=1-\langle v_\text{T}^2\rangle/(2\langle v_r^2\rangle)$
\citep{BT87}. Here, $\langle v_r^2\rangle$ and $\langle v_\text{T}^2
\rangle$ are the radial and the tangential velocity second moments.
For a wide range of cosmological simulations, they demonstrate
that the dark matter follows the equation of state $\beta\approx
1-1.15(1-\alpha/6)$ where $\alpha$ is the density power index (i.e.,
$\rho\sim r^{-\alpha}$). In the very center, dark matter halos are
roughly isotropic ($\beta\approx0$) with $\alpha\approx1$.
In the outer parts, violent
relaxation produces a density profile that asymptotically becomes
$\rho\sim r^{-4}$ \citep{Ja87} or $\rho\sim r^{-3}$ \citep*{NFW95},
for which the anisotropy parameter $\beta\approx0.5$ accordingly.

If violent relaxation proceeded to completion, then equipartition
would enforce equal kinetic energy in each direction and the velocity
distribution would be isotropic \citep{Ly67}. This appears to be the
case only at the centers of numerical simulations. Particles with
large apocenters respond only weakly to the fluctuating gravitational
field. Throughout most of the halo, this gives rise to an end point for
which the kinetic energy in the radial direction is roughly equal to
the sum of that in the two tangential directions. This seems to be
supported not only by the numerical simulations but also by the
observation of stars in elliptical galaxies \citep{vdM94}, whose
kinematics is also governed by the collisionless Boltzmann equation
with the gravitational potential. The purpose of this paper is to give
the DF of the dark matter which has this property. 

There has been much work on isotropic DFs \citep[see][]{BT87} of
gravitating systems. These are fine for the inner parts. On the other
hand, there has been much less work on DFs suitable for the radially
anisotropic outer parts of the dark matter halos. In particular, a
number of the suggestions in the literature for anisotropic DFs
\citep[e.g.,][]{Os79,Os79tr,Me85,Cu91} are unsuitable, as they yield
overwhelming radial anisotropy ($\beta\rightarrow1$) in the outer
parts, which is inconsistent with the simulations. While there exist
some suggestions on the form of anisotropic DFs with a more flexible
behavior of $\beta$ \citep[e.g.,][]{Ge91,Lo93}, recovering such DFs
for most density profiles is often analytically intractable
\citep[but see \citealp{De87} for a special case]{CL95}.

\section{Distribution functions with $\beta=1/2$}

The widely used ansatz for a DF of a spherical system with constant
anisotropy (parameterized by $\beta$) is
\beq
f(E,L)=L^{-2\beta}f_E(E)
\label{eq:ansatz}
\eeq
where $E=\psi-v^2/2$ is the binding energy per unit mass,
$L=rv_\text{T}$ is the specific angular momentum, and $\psi$ is the
relative potential. Integration of the DF over the velocity gives
%
\beq
\rho=r^{-2\beta}\,
\frac{(2\pi)^{3/2}\Gamma(1-\beta)}{2^\beta\Gamma(3/2-\beta)}
\int_0^\psi\!(\psi-E)^{1/2-\beta}f_E(E)\,dE.
\label{eq:den}
\eeq
The unknown function $f_E(E)$ then can be recovered from the integral
inversion formula \citep{Cu91,AE06};
\beq
f_E(E)=
\frac{2^\beta(2\pi)^{-3/2}}{\Gamma(1-\lambda)\Gamma(1-\beta)}\
\frac{d}{dE}\!\int_0^E\frac{d\psi}{(E-\psi)^\lambda}
\frac{d^nh}{d\psi^n}
\label{eq:disint}
\eeq
where $h=r^{2\beta}\rho$ is expressed as a function of $\psi$, and
$n=\lfloor(3/2-\beta)\rfloor$ and $\lambda=3/2-\beta-n$ are the
integer floor and the fractional part of $3/2-\beta$. This includes
Eddington's formula \citep{Ed16} for the isotropic DF as a special
case ($\beta=0$). The expression for the differential energy
distribution (DED) reduces to \citep[c.f.,][]{Cu91}
\beq
\frac{dM}{dE}=f_E(E)\,
\frac{(2\pi)^{5/2}\Gamma(1-\beta)}{2^{\beta-1}\Gamma(3/2-\beta)}
\int_0^{r_E}(\psi-E)^{1/2-\beta}r^{2(1-\beta)}dr.
\label{eq:ded}
\eeq
Here, $r_E$ is the radius of the largest orbit of a particle with
energy $E$, that is to say, $\psi(r_E)=E$.

If $\beta$ is a half-integer constant (i.e., $\beta=$ $1/2$, $-1/2$,
and so on), the expression for DF further reduces to
\beq
f(E,L)=\frac1{2\pi^2}\frac{L^{-2\beta}}{(-2\beta)!!}\
\left.\frac{d^{3/2-\beta}h}{d\psi^{3/2-\beta}}\right|_{\psi=E}.
\label{eq:dishalf}
\eeq
This involves only differentiations, as first noted by \citet{Cu91}.
For the simplest case of $\beta=1/2$, by utilizing the parameter
$r_E$, the expressions for the DF and the DED can be simply written
down as
\beq
\begin{split}&
f(E,L)=\frac{g(r_E)}{2\pi^2L}\,;\qquad
\frac{dM}{dE}=2\pi r_E^2g(r_E),
\label{eq:def}
\\&
g(r_E)=\left.\frac{\rho+r(d\rho/dr)}{(d\psi/dr)}\right|_{r=r_E}
=\frac{\rho r_E^2}{GM_r}\left.
\left(-1-\frac{d\ln\rho}{d\ln r}\right)\right|_{r=r_E},
\end{split}
\eeq
where $M_r$ is the enclosed mass within the sphere of radius of $r$.
\emph{So, the $\beta = 1/2$ case, which is desirable from the point of
view of the N-body simulations, is also very attractive
mathematically. That is, both the DF and DED can be found almost
trivially from the potential-density pair.}

\section{Cosmological halo models}

\subsection{Generalized NFW Profiles}

\begin{figure*}
\includegraphics[width=0.482\hsize]{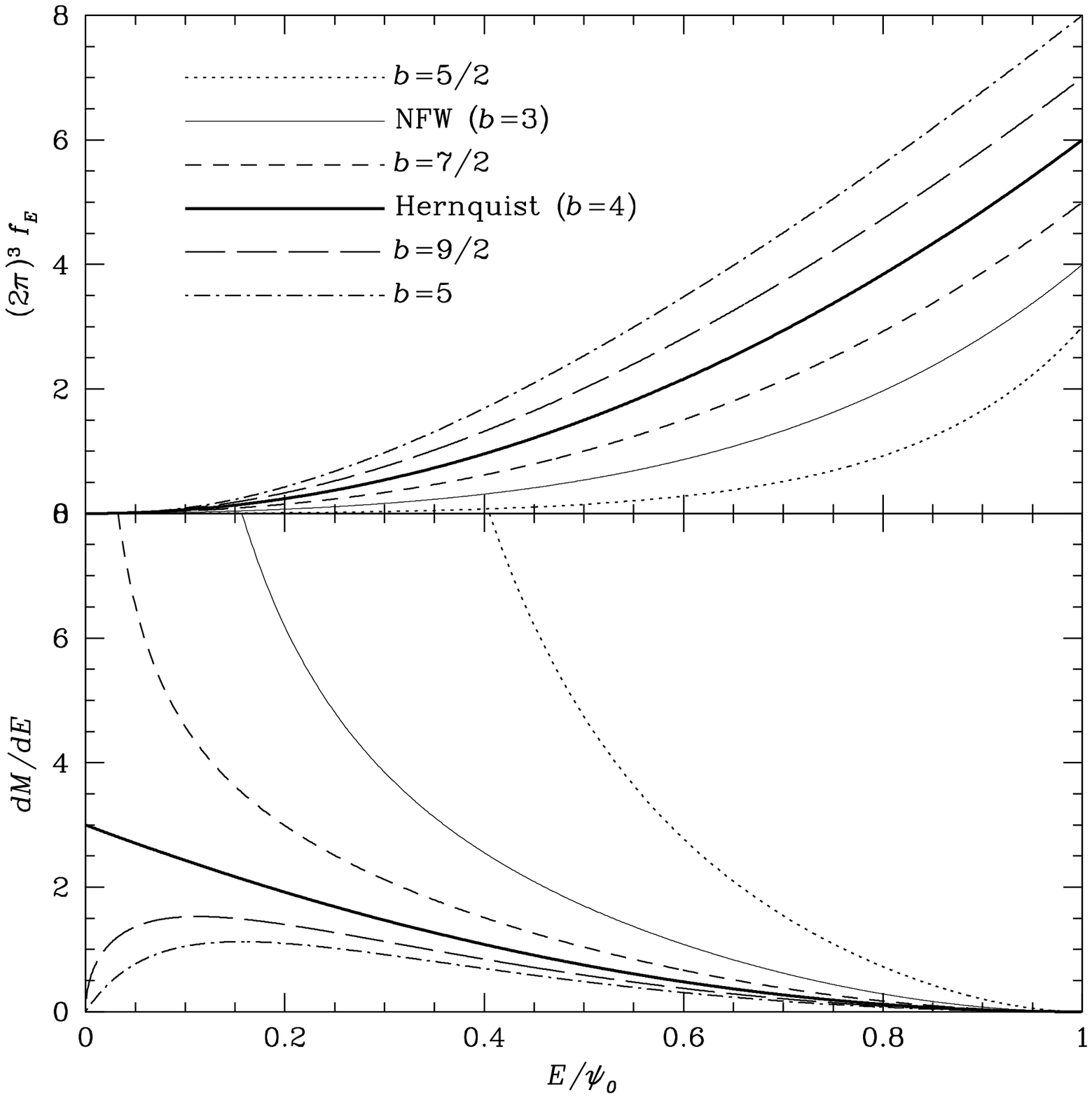}
\includegraphics[width=0.482\hsize]{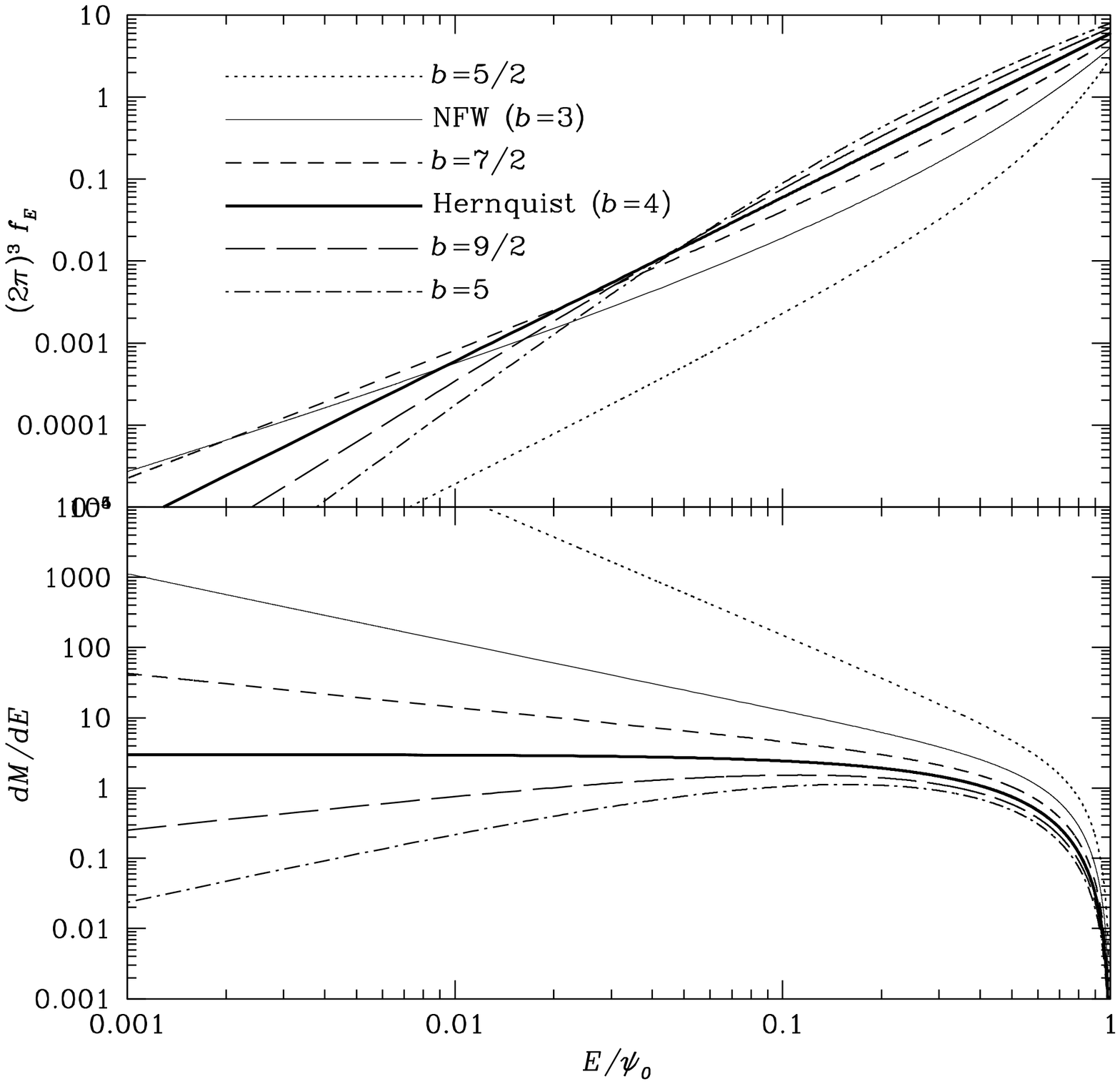}
\caption{\label{fig:bdf}\footnotesize
The energy part of the distribution function
(upper panel) and the differential energy distribution (lower panel)
of the Navarro-Frenk-White family with the constant anisotropy
parameter of $\beta=1/2$: dotted lines ($b=5/2$), thin solid lines
($b=3$; NFW profile), short-dashed lines ($b=7/2$), think solid lines
($b=4$; Hernquist model), long-dashed lines ($b=9/2$), dot-dashed
lines ($b=5$). Note that the models are normalized to the common
value of the depth of the central potential well, and, for $b\le3$,
the total mass is divergent.}
\end{figure*}

Let us consider a family of centrally cusped density profiles 
\beq
\rho=\frac{(b-2)\psi_0}{4\pi G}\frac{a^{b-2}}{r(r+a)^{b-1}}
\label{eq:nfw}
\eeq
where the parameter $b>2$ is the asymptotic density power index at
large radii, and $\psi_0$ is the depth of the central potential well.
Near the center, the density for every member of this family is always
cusped as $r^{-1}$. This reduces to the \citet{He90} model for $b=4$
whereas it becomes the `universal' halo density law or so-called
Navarro-Frenk-White (NFW)
profile \citep{NFW95} if $b=3$. The system has an infinite mass for
$2<b\le3$. On the other hand, if $b>3$, the finite total mass is given
by $GM_\infty= \psi_0a/(b-3)$.
The corresponding potential is
\bdm
\psi=\frac{\psi_0a}{r}\Ln{b-2}\left(\frac{r+a}{a}\right),
\edm
where, to reduce notational clutter, we have used the
`$q$-logarithm' function \citep{Ts88} defined to be
\bdm
\Ln{q}(x)\equiv\int_1^x\frac{dt}{t^q}=
\begin{cases}\
\left(x^{1-q}-1\right)/(1-q)&q\ne1\\\
\ln x&q=1\end{cases}.
\edm
Here, we note also a property of $q$-logarithm function, namely
$\Ln{q}(x^{-1})=-\Ln{2-q}(x)$. Its inverse is the `$q$-exponential'
function
\bdm
\Ex{q}(x)=\Ln{q}^{-1}(x)=
\begin{cases}\
\left[1+(1-q)x\right]^{1/(1-q)}&q\ne1\\\
\exp(x)&q=1\end{cases}.
\edm

The DF of the form of Eq.~(\ref{eq:ansatz}) can be found using
(with $G=\psi_0=a=1$)
\bdm
h=r^{2\beta}\rho=\frac{b-2}{4\pi}\frac{r^{2\beta-1}}{(1+r)^{b-1}}
\edm
and Eqs.~(\ref{eq:disint}) or (\ref{eq:dishalf}). However, for
$\beta=1/2$, the formulas~(\ref{eq:def}) enable us to write down the
DF for all the family using $r_E$ [here, $r_EE=\Ln{b-2}(1+r_E)$;
see Appendix~\ref{appendix} for $b=$ 7/3, 5/2, 8/3, 7/2, 4, or 5].
We find that
\begin{eqnarray}
f(E,L)&=&\frac{b-2}{(2\pi)^3L}\frac{(b-1)r_E}{(1+r_E)^bE-(1+r_E)^2}\,;
\label{eq:bdf}
\\
\frac{dM}{dE}&=&\frac{b-2}2\frac{(b-1)r_E^3}{(1+r_E)^bE-(1+r_E)^2}.
\label{eq:bde}
\end{eqnarray}
Here, $E$ ranges in the interval $[0,1]$ because $0\le E\le\psi\le1$.
Note that $r_E\rightarrow0$ as $E\rightarrow1$ and $r_E\rightarrow
\infty$ as $E\rightarrow0$. For all members of the family with $b>2$,
we find that the DF is non-negative for all accessible phase space
volume. The behavior of the energy part of the DF (Eq.~\ref{eq:bdf})
and the DED (Eq.~\ref{eq:bde}) of this family are shown in
Fig.~\ref{fig:bdf} for several values of $b$.


\citet{LM01} derived analytical expressions for various physical
properties of the NFW profile including the profiles of the velocity
dispersions and the kinetic and potential energy assuming isotropy,
constant anisotropy, or an Osipkov\citep{Os79,Os79tr}-Merritt\citep{Me85}
type DF.  However, they fell short of deriving any explicit DF with
the exception of the isotropic DF, for which they gave the result of
the numerical integration of the Eddington's formula. In fact, (the
numerical integrations of) the isotropic as well as the Osipkov-Merritt
DFs and DEDs for the NFW profile have been investigated in detail
by \citet{Wi00}. We nevertheless again note that the isotropic DF is
suitable only for the very inner parts of dark matter halos and that
the Osipkov-Merritt type DF is unsuitable in general, as it yields
overwhelming radial anisotropy ($\beta\rightarrow1$) in the outer
parts.

It is relatively straightforward to find the asymptotic behavior of
the DF and the DED near $E=1$ by means of Taylor series expansion of
$r_E$ at $E=1$. We find that $f(E=1,L)=2(b-1)/[(2\pi)^3L]$ tends to a
constant while $dM/dE\sim(1-E)^2\rightarrow0$ as $E\rightarrow1$. On
the other hand, the asymptotic behavior near $E=0$ for $b\ne3$ can be
derived from $(3-b)r_EE\approx r_E^{3-b}-1$ for $r_E\gg1$. Then, we
have
\bdm
r_E\sim
\begin{cases}\
E^{-1/(b-2)}&2<b<3\\\
E^{-1}&b>3
\end{cases},
\edm
for $0\le E\ll1$. Consequently, the asymptotic forms of the DF and the
DED are given by
\bdm
f_E\sim
\begin{cases}
E^{1/(b-2)}&2<b<3\\E^{b-2}&b>3
\end{cases};\
\frac{dM}{dE}\sim
\begin{cases}
E^{-1/(b-2)}&2<b<3\\E^{b-4}&b>3
\end{cases}
\edm
so that $f(E=0,L)=0$ whereas the DED diverges as $E\rightarrow0$ if
$b<4$ and is finite otherwise. In particular, $\lim_{E\rightarrow0}
(dM/dE)=0$ if $b>4$, and $dM/dE|_{E=0}=3$ if $b=4$.

For the NFW profile ($b=3$), the proper asymptotic form for the
inversion of $E=r_E^{-1}\ln(1+r_E)$ cannot be expressed using only
elementary functions. Nevertheless, the continuous nature
of the asymptotic behavior suggests that $dM/dE\sim E^{-1}$. In fact,
since $E\sim r_E^{-1}\ln r_E$ and $dM/dE\sim r_E/(r_EE-1)\sim r_E/\ln
r_E$ as $r_E\rightarrow\infty$, this is indeed the right behavior for
the NFW profile. We also note that it is possible to approximate as
$r_E\sim E^{-1}\Omega(E)$ and $f_E\sim E[\Omega(E)]^{-2}$ where
$\Omega(E)=-W_{-1}(-E)=\ln(E^{-1}\ln(E^{-1}\ln(E^{-1}\ldots)))
\approx\ln(E^{-1}\ln E^{-1})$. Here, $W_{-1}(x)$ is the real-valued
second branch of Lambert W-function \citep{W} such that
$W_{-1}(x)<-1$ for $-e^{-1}<x<0$. By comparison, we have
$f\sim E^{3/2}[\Omega(E)]^{-3}$ and $dM/dE\sim E^{-1}$ for the same
($E\rightarrow0$)-asymptotic behaviors of the isotropic DF and DED of
the NFW profile \citep{Wi00}.


\begin{figure}
\includegraphics[width=0.8\hsize]{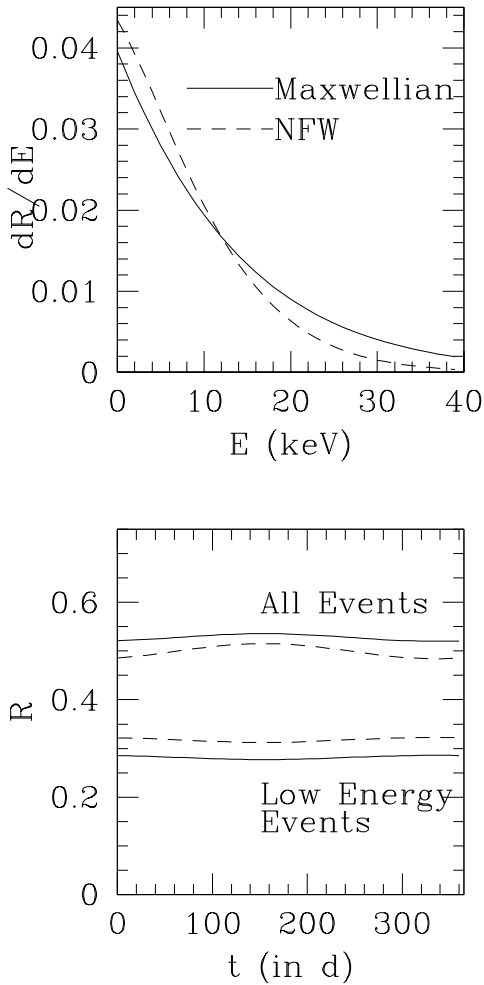}
\caption{\label{fig:dm}\footnotesize
Upper panel: The differential rate (in units
of events kg$^{-1}$ d$^{-1}$ keV$^{-1}$) for the case of 40 GeV dark
matter particles impinging on a cryogenic detector made of Ge.
Results are shown for a standard isothermal sphere with Maxwellian DF
(full line) and the NFW model with DF given by Eq.~(\ref{eq:bdf})
(dashed line). Lower panel: The annual modulation signal (in units of
events kg$^{-1}$ d$^{-1}$) for the isothermal sphere and NFW models.
The upper curves show the variation in the total rate, the lower
curves the variation in the low energy events ($<10\ \mbox{keV}$).}
\end{figure}

As a brief illustration of the application of our DF
(Eq.~\ref{eq:bdf}), we consider the direct detection of dark
matter. Such experiments work by measuring the recoil energy of a
nucleus in a low background laboratory detector that has undergone a
collision with a dark matter particle. Although the deposited energy
is tiny and the interactions are very rare, there are now many groups
searching for this effect worldwide \citep[e.g.,] [and references
therein]{SK05}. The detection rate depends on the masses $m_\chi$ and
$m_\text{N}$ of the dark matter particle and the target nucleus and
the elastic scattering cross-section $\sigma_0$ between them. It also
depends on the local dark matter density $\rho_0$ and the speed
distribution of the dark matter particles (in the rest frame of the
target).

The differential rate for detection per unit detector mass is given
by \citep[c.f.,][]{JKG96}
\beq
\left.\frac{dR}{dE}\right|_{E=\mathcal{E}}=
\frac{\sigma_0}{2m_\chi\mu^2}F^2(\mathcal{E})
\int\!\frac{d^3\bm{v}}{|\bm{v}|}\,f(E,L)\,
\Theta(|\bm{v}|-v_{\min}),
\label{eq:rate}
\eeq
where $\Theta(x)$ is the Heaviside unit step function, $\mathcal{E}$
is the recoil energy, $\mu^{-1}=m_\text{N}^{-1}+m_\chi^{-1}$ is the
reduced mass, and
\bdm
v_{\min}=\left(\frac{\mathcal{E}m_\text{N}}{2\mu^2}\right)^{1/2}.
\edm
In addition, $F(\mathcal{E})$ is the nuclear form factor, which is
commonly modeled at least for scalar interaction by \citep{Ah87,FFG88}
\bdm
F(\mathcal{E}) = \exp\left(-\frac{\mathcal{E}}{2\mathcal{E}_0}\right), 
\edm
where $\mathcal{E}_0$ is the nuclear coherence energy. Here, the DF is
normalized to the local density of the dark matter;
\bdm
\rho_0=\int\!d^3\bm{v}\,f(E,L).
\edm
Note the integral in Eq.~(\ref{eq:rate}) is over the velocity
with respect to the detector on Earth. Of course, the Earth revolves
around the Sun while the Sun moves with respect to the Galactic
inertial frame (in which the net angular momentum of the dark matter
halo vanishes). This produces an annual modulation in the signal,
which the experiments hope to detect.

The total event rate can be found by integrating over all detectable
energies. For the sake of definiteness, we consider a dark matter
particle with only scalar interactions and with a mass $m_\chi c^2=40\
\mbox{GeV}$ and cross-section $\sigma_0=4\times10^{-36}\
\mbox{cm$^{-2}$}$. The detector is made of $^{73}$Ge. The local halo
density is taken as $\rho_0c^2=0.3\ \mbox{GeV cm$^{-3}$}$
\citep{GGT95}.

We consider two models for the dark matter halo. The first is a
standard isothermal sphere, with a flat rotation curve of amplitude
$v_0=220\ \mbox{km s$^{-1}$}$. The DF is a Maxwell-Boltzmann
distribution \citep{BT87}. This is a useful benchmark, as the model is
widely used in dark matter studies \citep[e.g.,][]{JKG96}. The second
is a NFW model (Eq.~\ref{eq:nfw} with $b=3$) with $a=10\ \mbox{kpc}$,
normalized to provide the assumed local halo density. The DF is given
by Eq.~(\ref{eq:bdf}).

The results for the differential rate and the annual modulation signal
are shown in Fig.~\ref{fig:dm}. The total rate is lower by $\sim$10\%
for the NFW model compared against the isothermal sphere. But, the
peak-to-peak amplitude of the modulation signal has increased from
$\sim0.018\ \mbox{events kg$^{-1}$ d$^{-1}$}$ to $\sim0.030\
\mbox{events kg$^{-1}$ d$^{-1}$}$. This renders the dark matter
particle more detectable. However, this good news comes with a
caveat. If the experiment is only sensitive to low energy events
($\mathcal{E}<10\ \mbox{keV}$) then the peak-to-peak variation in the
modulation signal is actually smaller for the NFW model versus the
standard isothermal sphere. The main differences between the two
models is that the escape speed of dark matter particles is finite for
the NFW model, but infinite for the isothermal sphere. Therefore, the
former produces more low energy events, while the latter provides a
larger total number of events.

\subsection{The Gamma Spheres}

\begin{figure}
\includegraphics[width=\hsize]{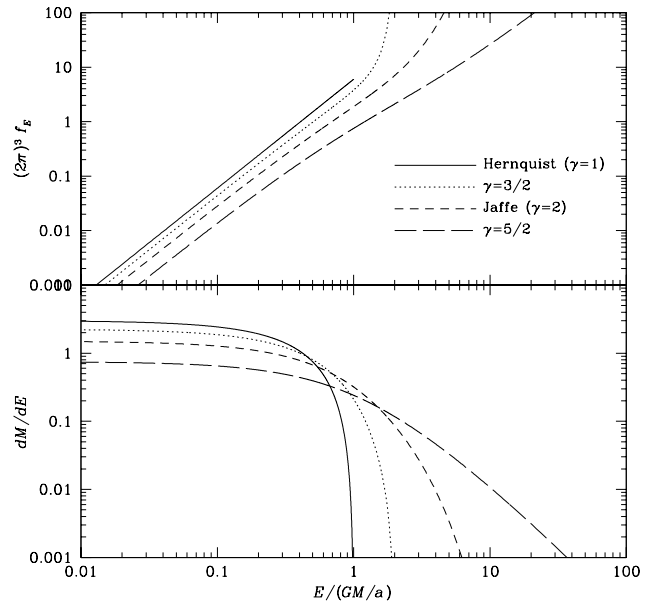}
\caption{\label{fig:gdf}\footnotesize
The energy part of the distribution
function (upper panel) and the differential energy distribution (lower
panel) of the $\gamma$ models with the constant anisotropy parameter
of $\beta=1/2$: solid lines ($\gamma=1$; Hernquist model), dotted
lines ($\gamma=3/2$), short-dashed lines ($\gamma=2$; Jaffe model),
long-dashed lines ($\gamma=5/2$). The models are normalized to the
same total mass. For $\gamma\ge2$, the central potential well depth is
infinite so that $E$ lies in the range $[0,\infty)$, whereas $0\le
E\le\psi_0=(2-\gamma)^{-1}(GM/a)$ for $\gamma<2$.}
\end{figure}

As a second example of our formulas~(\ref{eq:def}), we consider the
potential-density pair of the $\gamma$ model \citep{De93,Tr94}
\bdm
\rho=\frac{(3-\gamma)M}{4\pi}\frac a{r^\gamma(r+a)^{4-\gamma}}
\,;\quad
\psi=\frac{GM}a\Ln{3-\gamma}\left(\frac{r+a}r\right).
\edm
Here, the parameter $\gamma<3$ is the three-dimensional central
density slope (i.e., the central density is cusped as $r^{-\gamma}$ if
$\gamma>0$), and $M$ is the total mass. At large radii, the density
falls off as $r^{-4}$ for every member of the family. The central
potential well depth is infinite if $2\le\gamma<3$, whereas, if
$\gamma<2$, the potential is bounded as
$0\le\psi\le(2-\gamma)^{-1}(GM/a)$. The family contains the
\citet{He90} model ($\gamma=1$) and the \citet{Ja83} model
($\gamma=2$) as special cases.

If we define $y=r/(r+a)$, it is easy to write down ($G=M=a=1$)
\bdm
h=r^{2\beta}\rho=
\frac{3-\gamma}{4\pi}\frac{(1-y)^{4-2\beta}}{y^{\gamma-2\beta}}.
\edm
Here, $h$ can be written explicitly as a function of $\psi$ using
$y=\Ex{\gamma-1}(-\psi)$. Then, for $\beta=1/2$, from
Eq.~(\ref{eq:dishalf}), we find the DF and the DED,
\begin{eqnarray}
f(E,L)&=&\frac{3-\gamma}{(2\pi)^3L}
\frac{(1-y_E)^2}{y_E}\left[(4-\gamma)y_E+(\gamma-1)\right]\,;
\label{eq:gdf}
\\
\frac{dM}{dE}&=&
\frac{3-\gamma}2\left[(4-\gamma)y_E+(\gamma-1)\right]y_E
\label{eq:gde}
\end{eqnarray}
where $y_E=r_E/(1+r_E)=\Ex{\gamma-1}(-E)$, which can be always
expressible using elementary functions of $E$. Since $0\le y_E\le1$,
the DF is everywhere non-negative only if $1\le\gamma<3$ \citep[c.f.,]{AE05}.
The behavior of the energy part of the DF (Eq.~\ref{eq:gdf}) and the
DED (Eq.~\ref{eq:gde}) of the $\gamma$ models are shown in
Fig.~\ref{fig:gdf} for several values of $\gamma$.

As $E\rightarrow0$ ($y_E\rightarrow1$), we find that $f_E(E)\sim
E^2\rightarrow 0$ whereas the DED is always finite with the limiting
value of $3(3-\gamma)/2$. On the other hand, the asymptotic behavior
as $E\rightarrow\psi_0$ ($y_E\rightarrow0$), where $\psi_0=\infty$ for
$\gamma\ge2$ or $\psi_0=(2-\gamma)^{-1}$ for $\gamma<2$, are found to
be $f_E\sim y_E^{-1}$ and $(dM/dE)\sim y_E$ where
\bdm
y_E\sim
\begin{cases}\
E^{-1/(\gamma-2)}&2<\gamma<3\\\
e^{-E}&\gamma=2\\\
(\psi_0-E)^{1/(2-\gamma)}&1<\gamma<2
\end{cases},
\edm
except for the limiting case of $\gamma=1$ (the Hernquist model), for
which $f(E,L)=3(4\pi^3)^{-1}L^{-1}(1-y_E)^2=3(4\pi^3)^{-1}E^2L^{-1}$
and $(dM/dE)=3y_E=3(1-E)^2$. Note that the asymptotic behavior of the
DED at both limits are the same as those for the isotropic DF
\citep{De93} for $1<\gamma<3$ despite the fact that the behavior of the
DFs are rather distinct.

\section{The Universal Asymptotic Behavior}

The asymptotic behaviors derived for the DF and DED in the preceding
sections strongly suggest that they are simply determined by the
anisotropy parameter and the density power index at the center and at
the infinity. Even more interestingly, the DED appears to be
completely determined (up to scale) by the density power index alone.
Assuming that this is indeed the case, we predict asymptotic behaviors
of generic DFs and DEDs by generalizing the method of \citet{HM91},
while allowing for anisotropy by means of the ansatz~(\ref{eq:ansatz}).

First, the self-consistent potential due to the asymptotic density
profile of $\rho\sim r^{-\alpha}$ is
\bdm
\begin{split}&
\psi\sim
\begin{cases}
r^{-1}&\alpha>3\\
r^{-(\alpha-2)}&2<\alpha<3
\end{cases}\\&
(\psi_0-\psi)\sim r^{2-\alpha}\qquad\alpha<2
\end{split}.
\edm
Here, since the case that $\alpha>3$ is only allowed for the
asymptotic falloff at infinity, the potential should tend to the
Keplerian finite mass limit. Next, if we assume $f_E$ to be roughly
scale-free with $f_E\sim E^n$ (for $\alpha>2$) or
$f_E\sim(\psi_0-E)^n$ (for $\alpha<2$),
we find from Eq.~(\ref{eq:den})
\bdm
\rho\sim
\begin{cases}
r^{-2\beta}\psi^{n-\beta+3/2}&\alpha>2\\
r^{-2\beta}(\psi_0-\psi)^{n-\beta+3/2}&\alpha<2
\end{cases}.
\edm
In order for this to be $\rho\sim r^{-\alpha}$ with the
self-consistent potential, we should have
\beq
L^{2\beta}f(E,L)\sim
\begin{cases}
E^{\alpha-\beta-(3/2)}&\alpha>3\\
E^{\beta-(1/2)+p}&2<\alpha<3\\
\exp[2(1-\beta)E]&\alpha=2\\
(\psi_0-E)^{\beta-(1/2)-(-p)}&2\beta<\alpha<2
\end{cases}
\label{eq:asydf}
\eeq
where $p=2(1-\beta)/(\alpha-2)$. Here, $\alpha$ and $\beta$ are the
limiting values at the center (for behavior near $E\sim\psi_0$) or the
asymptotic value toward infinity (for behavior near $E\sim0$). As for
the DED, by changing the integration variable to $\psi/E$ (for
$\alpha>2$) or $(\psi_0-\psi)/ (\psi_0-E)$ (for $\alpha<2$) in
Eq.~(\ref{eq:ded}), and combined with Eqs.~(\ref{eq:asydf}),
we find that,
\beq
\frac{dM}{dE}\sim
\begin{cases}
E^{\alpha-4}&\alpha>3\\E^{-1/(\alpha-2)}&2<\alpha<3\\
\exp(-E)&\alpha=2\\(\psi_0-E)^{1/(2-\alpha)}&2\beta<\alpha<2
\end{cases}.
\label{eq:asyded}
\eeq
It is independent of the anisotropy. We note that the coefficient for
the leading order term of $f_E$ changes its sign at $\alpha=2\beta$ so
that the result is invalid at the limit $\alpha=2\beta$ and that the
DF is unphysical for $\alpha<2\beta$ \citep{AE05}.

Here, since $\alpha<3$ at the center, $\lim_{E\rightarrow\psi_0}
dM/dE=0$ (where $\psi_0=\infty$ if $2\le\alpha<3$) for
all physical values of $\alpha$. On the other hand, the behavior of
$dM/dE$ near $E=0$ for a finite mass system (i.e., $\alpha>3$) implies
that $dM/dE$ diverges for $\alpha<4$ while it is finite for $\alpha\ge
4$ (in particular, $dM/dE\rightarrow0$ if $\alpha>4$). It has been
argued before that violent relaxation produces an $r^{-4}$ density
falloff at large radii \citep{Ja87}. This may be inferred from the
generic behavior of $dM/dE$ at the asymptotic limit $E\rightarrow0$
(note that the behavior of $dM/dE$ near $E=0$ is dominated by the
particles at large radii). That is, the loss of loosely bound particles at
large radii due to velocity perturbations is much more significant if
the initial density falloff is shallower than $r^{-4}$ while it
becomes rather insignificant once the density falloff gets steeper
than $r^{-4}$. Therefore, any perturbation drives systems with
initially shallower density falloff to settle toward the $r^{-4}$
falloff or a slightly steeper slope. Note that this argument is
completely independent of the anisotropy since the asymptotic form of
$dM/dE$ is also independent of $\beta$.

If one considers the case that the gravitational field is dominated
not by the self-consistent potential but by the Keplerian field due to
the central black hole, it is straightforward to see that the same
argument leads to the asymptotic behavior of $f_E\sim
E^{\alpha-\beta-3/2}$ and $dM/dE\sim E^{\alpha-4}$ for all allowed
values of $\alpha>\beta+1/2$ except also for the limiting case
$\alpha=\beta+1/2$, for which the otherwise leading term identically
vanishes.

\section{Conclusions}

Except close to the very center, dark matter halos have distribution
functions (DFs) that are radially anisotropic with an anisotropy
parameter $\beta\approx0.5$. Constant anisotropy distribution
functions with $\beta=1/2$ are very simple to construct -- far simpler
than Eddington's awkward Abel transformation pair for the isotropic model.
This paper provides simple inversion formulas~(\ref{eq:def}),
which enables such DFs and the corresponding
differential energy distributions to be built for any spherical model,
provided that its potential and density are known.

We have also given the \emph{generic} form of the DF
for any spherical model with a local density power index $\alpha$
($\rho\sim r^{-\alpha}$) and anisotropy parameter $\beta$. In the
central parts of simulated halos where $\alpha\approx1$ and
$\beta\approx0$, we find $f(E)\sim(\psi_0-E)^{-5/2}$ and
$dM/dE\sim(\psi_0-E)$. In the envelopes of simulated halos with
$\beta\approx1/2$, we have DFs, which range from
\bdm
f(E,L)\sim\frac{1}{L}\frac{E}{\Lambda(E)}\,;\qquad\frac{dM}{dE}\sim\frac{1}{E}
\edm
if $\alpha\approx3$ [where $\Lambda(E)$ is some function that diverges
or vanishes no faster than finite power of the logarithm as
$E\rightarrow0$], to
\bdm f(E,L)\sim\frac{E^2}{L}\,;\qquad\frac{dM}{dE}\sim E^0
\edm
if $\alpha\approx4$. Finally, we have argued that falloff of the
density at large radii must have evolved to $\rho\sim r^{-4}$ or
steeper in the long run. This argument is independent of the
velocity anisotropy.

Using our DF for the NFW profile, we calculated the direct detection
rate for a cosmological halo model with a radially anisotropic DF. We
showed that the annual modulation signal is larger in radially
anisotropic ($\beta=1/2$) cosmological halo models than in isotropic
isothermal spheres. This may be welcome good news for dark matter
experimentalists.

\appendix
\section{}\label{appendix}

If $b=$ 7/3, 5/2, 8/3, 7/2, 4, or 5, then the integral of motion $r_E$
for the generalized NFW profiles is expressible analytically in terms
of the energy by solving a quadratic or linear equation, viz
($G=\psi_0=a=1$)
\begin{widetext}
\[
\begin{split}
r_E=\frac{4(1-E)}{E^2}\qquad\mbox{for }\ b=\frac52\,;&\qquad
r_E=\frac{1-E}E\qquad\mbox{for }\ b=4\,;\\
r_E=\frac{9(3-4E^2)+3^{3/2}(3-2E)^{3/2}\sqrt{1+2E}}{16E^3}\qquad
\mbox{for }\ b=\frac73\,;&\qquad
r_E=\frac{3^{3/2}\sqrt{4-E}}{2E^{3/2}}-\frac9{2E}\qquad
\mbox{for }\ b=\frac83\,;\\
r_E=\frac{4-E-E^{1/2}\sqrt{8+E}}{2E}\qquad
\mbox{for }\ b=\frac72\,;&\qquad
r_E=\frac{1-4E+\sqrt{1+8E}}{4E}\qquad
\mbox{for }\ b=5.
\end{split}
\]
\end{widetext}

For all other values of $b$, the function $r_E(E)$ is straightforward
to construct numerically.

\end{document}